\documentclass{vldb}

\usepackage{balance}  
\usepackage{amsmath}
\usepackage{mathtools}
\usepackage{hyperref}             
\usepackage{listings}
\usepackage{color}
\usepackage{arydshln}
\usepackage{tikz-cd}
\usepackage{microtype}
\usepackage{paralist}
\usepackage{graphicx}

\newtheorem{theorem}{Theorem}[section]
\newtheorem{lemma}[theorem]{Lemma}

\newcommand{\m}[1]{\mbox{\it #1}}

\begin{document}


\title{A View-based Programmable Architecture\\ for Controlling and Integrating Decentralized Data}
\subtitle{[Vision Paper]}



%
%
%
%

\numberofauthors{1} 

\author{
%
%
\alignauthor
Yasuhito~Asano$^3$,
Soichiro~Hidaka$^4$,
Zhenjiang~Hu$^1$,
Yasunori~Ishihara$^5$,
Hiroyuki~Kato$^1$,
Hsiang-Shang~Ko$^1$,
Keisuke~Nakano$^6$,
Makoto~Onizuka$^2$,
Yuya~Sasaki$^2$,
Toshiyuki~Shimizu$^3$,
Kanae~Tsushima$^1$,
Masatoshi~Yoshikawa$^3$\\
\smallskip
       \affaddr{$^1$National Institute of Informatics, $^2$Osaka University, $^3$Kyoto University,}\\
       \affaddr{$^4$Hosei University, $^5$Nanzan University, $^6$The University of Electro-Communications}\\
}

\maketitle

\begin{abstract}
The view and the view update are known mechanism for controlling access of data and for integrating data of different schemas.
Despite intensive and long research on them in both the database community and the programming language community, we are facing difficulties to use them in practice. The main reason is that we are lacking of control over the view update strategy to deal with inherited ambiguity of view update for a given view.

This vision paper aims to provide a new language-based approach to controlling and integrating decentralized data based on the view, and establish a software foundation for systematic construction of such data management systems. Our key observation
is that
\emph{a view should be defined through a view update strategy rather than a query}.
In other words,
the view definition should be extracted from the view update strategy, which is in sharp contrast to the traditional approaches where the view update strategy is derived from the view definition.

In this paper, we present the first programmable architecture with a declarative language for specifying update strategies over views, whose unique view definition can be automatically derived, and show how it can be effectively used to control data access, integrate data generally allowing coexistence of GAV (global as view) and LAV (local as view), and perform both analysis and updates on the integrated data. 
We demonstrate its usefulness through development of a privacy-preserving ride-sharing alliance system, discuss its application scope, and highlight future challenges.
\end{abstract}

\section{Introduction}

Along with the continuous evolvement of data management systems for the new market requirements, centralized systems, which had often produced huge and monolithic databases, have been replaced by decentralized systems in which data are maintained in different sites with autonomous storage and computation capabilities. The owner of the data stored on a site may choose to show what information should be exposed and how its information should be updated by other systems. On the other hand, the systems would like to integrate data from different sites and perform analysis and update on the integrated data. The goal of this vision paper is to combine the advanced technologies developed in both the database community and the programming language community to establish software foundations to control and integrate these distributed decentralized data.

\subsubsection*{View update problem in DB}


View plays an important role in controlling access of data \cite{FeSW81,Foster:2009:USV:1602936.1603620} and for integrating data of different schemas \cite{DoHI12,GHMT18}, since it was first introduced by Codd about four decades ago \cite{Codd74}.
It is a relation derived from base relations, which is helpful to describe dependencies between relations and achieve database security within an authorization framework.

Deeply associated with view is the classic {\em view update problem} \cite{Bancilhon:81,Dayal:82,Masu84,Larson:1991,Hegner:04}: given a view defined by a query over base relations, show how to systematically reflect the changes made to the view as updates to the original base relations. Put it more concretely,
given that $s$ represents a database state, $Q$ is a view definition, $Q(s)$ represents the view state from $s$, and $u$ represents the update operation issued to $Q(s)$, the view update problem is defined as finding a translation $T$ of $u$ such that the following commutative diagram holds.
\[
\Large
\begin{tikzcd}
  s \arrow[r, "Q"] \arrow[d, "T(u)"]
    & v \arrow[d, "u"] \\
  s' \arrow[r, "Q"]
    &v'
\end{tikzcd}
\]

Despite a long and intensive study \cite{Keller:1986,Masunaga:2017:IAU:3022227.3022239} of the view updating in the database community, as discussed in \cite{Masunaga:2017:IAU:3022227.3022239}, there are few really practical systems that can fully support view updating.
It is essentially impossible to obtain a unique solution to a view update, because of potentially many incomparable strategies to reflect a view update.

This calls for a general method to solve a fundamental tension
between expressiveness and realizability in the view update problem.
The richer language we use for defining views, the more difficult it becomes to find corresponding functions to reflect the updates on the view to that on the base relations.

\subsubsection*{Bidirectional transformation (BX) in PL}

To deal with this tension, many researchers in the programming language community have been attracted to generalize the concept of the view update problem to be a general synchronization problem \cite{GRACE:09,Terwilliger:2012}, and designed various domain specific languages \cite{Lenses,XLHZ07,Bohannon:08,Barbosa:2010,Hidaka:10} to support so-called {\em bidirectional transformation}, which generalizes the manipulation of data from relations to other data types, and allows views to be materialized.

A bidirectional transformation (BX) consists of a pair of
transformations:
\[
\Large
\begin{tikzcd}
  s \arrow[r, "get"] \arrow[d, bend left = 80]
    &  v \arrow[d] \\
  s'
    & \arrow[l, "put"] v'
\end{tikzcd}
\]
Here, the {\em forward} transformation $get(s)$ is used to
produce a target view \m{v} from a source \m{s}, while the {\em putback}
transformation $put(s,v)$ is used to reflect updates on the view
$v$ to the source $s$.
These two transformations should be {\em
well-behaved} in the sense that they satisfy the following
round-tripping laws.
\[
\begin{array}{lllr}
put(s, get(s))  &=& s \qquad & \textsc{GetPut} \\
get(put(s, v)) &=& v & \textsc{PutGet}
\end{array}
\]
The \textsc{GetPut} property requires that no changing on the view shall
be reflected as no changing on the source, while the \textsc{PutGet}
property requires all changes in the view to be completely reflected
to the source so that the changed view can be computed again by
applying the forward transformation to the updated source.
Exact correspondence between the notion of well-behavedness in BX and the properties on view updates such as translation of those under a constant complement \cite{Bancilhon:81,Dayal:82}, has been extensively studied in \cite{BeSm03}.

It has been demonstrated in \cite{Bohannon:06} that this language-based approach is useful to help solving the view update problem with
a bidirectional query language, in which every expression can be interpreted forwardly
as a view definition and backwardly as an update strategy.

One appealing feature of this language-based approach is its powerful type system, which includes record-level predicates and functional dependencies and can fully guarantee that update strategies are well-behaved. However, this solution is not that satisfactory, because it still cannot solve the issues of ambiguity of update strategies for a given view definition.

\subsubsection*{Problem: lack of effective control of update strategy}

The main difficulty in using these techniques to control and integrate distributed decentralized data lies in the inherited ambiguity of the update strategy for a given query or a forward transformation.
The problem is that we are lacking of effective way of controlling over the update strategy (or the putback transformation); it would be awkward and counterintuitive to obtain our intended update strategy by changing the view definition that is under our control, when the view definition becomes complicated.

We have been taken it for granted that a view should be defined by a query and that a sound and intended update strategy should be automatically derived even if it is known that automatic derivation of an intended update strategy is generally impossible \cite{Keller:1986}. Now it is time to consider seriously the following two fundamental questions: (1) Must views be defined by queries? and (2)
Must update strategies be automatically derived?

\subsubsection*{Our vision: a programmable architecture}

This vision paper aims to provide a new language-based approach to controlling and integrating decentralized data based on the view, and establish a software foundation for systematic construction of such data management systems. Our key observation is:
\begin{quote}
{\em A view should be defined through a view update strategy to the base relations rather than a query  from them}.
\end{quote}
This new perspective is in sharp contrast to the traditional approaches, and it actually gives an answer to the above two questions: a view is not necessary to be defined as a query, and an update strategy with human insight should be definable.

This vision stems from the recent work on the putback-based approach \cite{HuPS14,FiHP15,KoZH16,KoHu18} to bidirectional programming. The key idea is that although there are many $put$s that can correspond to a given $get$, there is at most one $get$ that can correspond to a given $put$, and such $get$ can be derived from $put$. In other words in terms of view and view update, we have that
\begin{quote}
  {\em for a view definition and a view update strategy, while
there may be many view update strategies for a given view definition, there is a unique view definition (if it exists) that corresponds to a view update strategy and this view definition can be derived.}
\end{quote}
This new perspective on view implies that we should design a language for describing view update strategies and treat the view definition as side-effect of the view update strategy. Following this line, we have designed BiGUL \cite{KoZH16, KoHu18}, a tiny  putback-based BX language to support programming putback functions declaratively while automatically deriving the corresponding unique forward transformation. It is interesting to investigate how to extend BiGUL to describe update strategies on relations.

Our main technical contributions can be summarized as follows.

\begin{itemize}
\item We present the first language for specifying update strategies over views on base relations, whose unique view definition can be automatically derived. We demonstrate how it is effectively used to control data access, to integrate data generally allowing coexistence of GAV (global as view) and LAV (local as view) \cite{DoHI12}, and to perform both analysis and updates on the integrated data.

\item We propose a novel view-based software architecture for systematic construction of a management system for controlling and integrating decentralized data. We highlight how this higher-level architecture can be implemented with PostgreSQL, where updates can be incrementally propagated between the view and the base relations and the well-behavedness in the higher-level architecture can be well preserved.

\item We demonstrate and validate this new approach through development of an application of a ride-sharing alliance system, where we can systematically obtain a robust implementation of the system in PostgreSQL, based on our view-based programmable architecture. The prototype implementation is available online\footnote{https://github.com/hiroyukikato/DataIntegration}.

\end{itemize}


The organization of the rest of the paper is as follows. We start with an overview of putback-based BX, the underlying foundation of this paper, in Section \ref{sec:background}, and give a motivation example of a ride-sharing alliance system in Section \ref{sec:ridesharing}. We then propose our view-based programmable architecture in Section \ref{sec:architecture}, present our putback-based language for specifying view update strategies in Section \ref{sec:language}, and discuss its implementation in Section \ref{sec:implementation}. We discuss the application scope, challenges, and the evaluation criteria in Section \ref{sec:discussion}, and give remarks on related work in Section \ref{sec:relatedWork}. Finally, we conclude the paper in Section \ref{sec:conclusion}.

\section{Foundation: Putback-based BX}
\label{sec:background}
\label{sec:putbx}

As discussed in the introduction, lots of work \cite{Lenses,Bohannon:06,Bohannon:08,Hofmann:2011,XLHZ07,MHNHT07,Voigt09,Hidaka:10} on BX has been devoted to the {\em get-based}
approach, allowing users to write the forward
transformation \m{get} and deriving a suitable putback transformation.
While the get-based approach is friendly,
a \m{get} function may not be injective, so there may exist
many possible  functions that can be combined with it to form a
BX and there is no way to control the choice of \m{put} through
the change of \m{get}.
This ambiguity of \m{put} is what makes bidirectional
programming challenging and unpredictable in practice.

In contrast to the get-based approach, the putback-based approach allows users to write the backward
transformation \m{put} and derives a suitable \m{get} that can be
paired with \m{put} to form a bidirectional transformation if it exists.
Interestingly, while \m{get} usually loses information
when mapping from a source to a view, \m{put} must preserve information
when putting back from the view to the source, according to the
\textsc{PutGet} property.

In the following, we recap the two important facts in \cite{HuPS14}, showing that "putback"
is the essence of bidirectional programming.
The first fact is that, for a \m{put}, there exists at most one \m{get}
that can form a BX with it. This is in sharp contrast to get-based
bidirectional programming, where many \m{put}s may be paired with a \m{get}
to form a BX.

\begin{lemma}[Uniqueness of \m{get}]
\label{lemma:injective}
Given a \m{put} function, there exists at most
one \m{get} function that forms a well-behaved BX.
\end{lemma}

The second fact is that it is possible to
check \emph{validity} of \m{put} in the sense that there is a $get$ that can be paired with $put$ to form a BX.
The following are two important properties on \m{put}.
\begin{itemize}
\item
 The first, that we call \emph{view determination}, says that equivalence
of updated sources produced by a \m{put} implies equivalence of views that are put back.
\begin{align*}
	\label{PutDet}
	\tag*{\textsc{ViewDetermination}}
	\forall~s,s',v,v'.~put(s,v)~=~put(s',v')~\Rightarrow~v~=~v'
\end{align*}

\item The second, that we call \emph{source stability}, denotes a slightly stronger notion of surjectivity for every source:
\begin{align*}
	\label{PutStable}
	\tag*{\textsc{SourceStability}}
	\forall~s.~\exists~v.~put(s,v)~=~s
\end{align*}
\end{itemize}
Actually, these two properties together provide an equivalent characterization of
the validity of \m{put}.
\begin{theorem}
\label{th:put2}
A \m{put} function is valid if and only if it satisfies the \ref{PutDet} and \ref{PutStable} properties.
\end{theorem}

BiGUL \cite{KoZH16,KoHu18} is a tiny
putback-based bidirectional language,
which grew out of the work \cite{PaHF14,PaZH14}.
In this paper, we will design a new bidirectional relational update language based on the idea of BiGUL.

\section{Running Example}
\label{sec:ridesharing}

To explain our programmable architecture and our implementation concretely,
we shall consider an example of a ``privacy-preserving ride-sharing alliance system''.
Being simple, this example gives a good demonstration of the need for controlling and integrating decentralized data.


Ride-sharing has become popular as an application which allows a person other than professional taxi drivers to provide a car service using his/her privately owned vehicle.
As companies who want to enter into the ride-sharing market increase, it is expected that ``alliances'' between companies also increase.
A ride-sharing alliance system receives requests from passengers and matches each request to a vehicle belonging to one of the companies.
In this system, companies might obtain more chances to have beneficial passengers, while passengers might have more choices of companies.
This system might consist of ride-sharing companies and an mediator (a third party, trusted in some degree) who integrates and analyzes the vehicle data of the companies.
The following is one of the possible scenarios.
\begin{enumerate}
\item A passenger sends a request to the ride-sharing alliance system (mediator) to book a taxi. \vspace{-2mm}
\item The mediator analyzes the user request and shows a candidate list of $K$ taxis to the passenger.\vspace{-2mm}
\item The passenger chooses a taxi and attempts to book the taxi.\vspace{-2mm}
\item The mediator sends the update request to the local database of a company that maintains the selected taxi.\vspace{-2mm}
\item If the company accepts the update, the company sends the ack of SUCCESS to the mediator.\vspace{-2mm}
\item Otherwise, the company sends the ack of FAIL. Back to Step 2.
\end{enumerate}

Meantime, it is important to control/protect the privacy of passengers and drivers in ride-sharing.
Because drivers are not professionals, passengers might not disclose their important locations to many drivers, and vice versa.
In the ride-sharing alliance system, ``privacy-preserving'' for passengers and drivers should mean to reduce the number of companies which know the precise location of a passenger and the number of vehicles whose precise locations are known to the mediator.
Previous researches about privacy-preserving ride-sharing \cite{aivodji2016privacy,goel2016privacy,tong2017privacy} have not considered such a system.
We present a solution in this paper as a demonstration of our proposal.


Let us consider the requirements for the privacy-preserving ride-sharing alliance system.
To reduce the vehicles whose precise locations are known to the mediator, the mediator should recommend vehicles for each request by integrating data of approximate areas of vehicles.
In addition, the system has to allow each company to control its own privacy policy for disclosing their vehicle data and own data update strategy for accepting requests, while keeping the consistency between the approximate area data of the mediator and the precise location data of each company.
Therefore, a desired architecture for the system should enable each company to easily describe its own privacy policy and update strategy.
Once if the policy and strategy are described, the architecture has to guarantee mediator's data and companies' data to be updated automatically keeping the consistency between the approximate area data of the mediator and the precise location data of each company.
We will discuss below that our view-based programmable architecture provides a solution for the privacy-preserving ride-sharing alliance system.


%

\section{A Programmable Architecture}
\label{sec:architecture}

%
This section gives an overview of our programmable architecture with an abstract explanation,
and shows how each part of the architecture is programmed
using a ride-sharing alliance system as an example.


\begin{figure}
  \centering
  \includegraphics[width=.98\columnwidth]{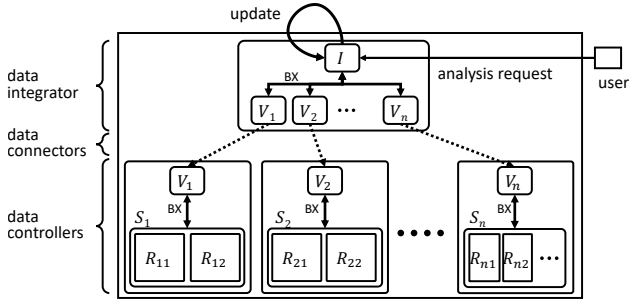}
  \caption{Overview of Programmable Architecture}
  \label{fig:architecture}
\end{figure}

Figure \ref{fig:architecture} shows the architecture
of our programmable model for controlling and integrating decentralized data.
%
%
It basically consists of three parts.
\begin{compactdesc}
\item[data controller] The lower part represents data sources
that provide data and accept updates in a controlled manner.
Each data source consists of pairs of data source $(S_i)$ and
its view $(V_i)$. BX between them,
 expressed in our update language,
is used to control how the
source is exported as its view, as well as how the updates to the
view are reflected to the source.

\item[data integrator] The upper part represents a mediator for data integration,
analysis, update, and update propagation.
$I$ represents the integrated view over the views $V_1,\ldots, V_n$
exported from data controllers. The BX between the integrated view
and each view of the controllers is also expressed in our update language.
\item[data connector] A connection between the data controllers the data integrator
(the dotted arrows in the figure).
\end{compactdesc}
The core parts in this programmable architecture are two BXs that are used in the data controller and in the data integrator respectively. They play an important role in controlling and integrating the data. 
Note that the analysis and update that are conducted on the integrated view in the integrator is common in most database management systems, which analyze the data and update them accordingly.

To illustrate, we show below how these two BXs are programmed for our running example, but leave the details of our update language for describing these BXs in Section \ref{sec:language}.

%

In the privacy-preserving ride-sharing alliance system, we first consider controlling data
in each taxi company.
Each taxi company as a data controller has its own database as $S_i$, from which
only a small portion ($\mathit{V}_i$) is exported to the mediator for queries and updates,
where updates are directly programmed as
controlled data sharing (Section~\ref{sec:RU-sharing}).
For example, updates to $V_i$ are redirected to each relation ($R_{ij}$)
through the language constructs to split relations vertically (column-wise)
or horizontally (row-wise). Note that in both cases, queries to generate
$V_i$ from $S_i$ are automatically derived from such programs 
(Section~\ref{sec:RU-query-derivation}).

Next, we consider integration of decentralized data.
Our update language for programming BX is able to program selective acceptance of the updates.
Suppose the update strategy of the first company does not allow updates to
area data of their vehicles.
If that attribute is changed on $V_1$ (at the data controller), then
the putback program rejects such updates
.
Such views ($V_1, \ldots,V_n$) are used for the mediator (as a data integrator) to form an integrated view that is used
for requests for booking/picking-up taxis by the passengers (``users'' in Figure~\ref{fig:architecture}).
Updates to the integrated view may trigger the updates on each view of the taxi
company. Such propagation can be programmed 
as in
Section~\ref{sec:RU-integration}, to route updates using company's ID.
The updates are sent to the views of each company through the connectors.
Some of such updates may not be acceptable, and such update strategies are
programmed as mentioned above.

It is worth noting that our update language allows GAV and LAV to coexist seamlessly
in this architecture, though our views are materialized while views in GAV and LAV
are usually virtual. GAV corresponds to the query derived from the
program in the data integrator because the derived query combines
$V_1, \ldots, V_n$ to create the integrated view $I$.
On the other hand, LAV in general is to create local databases from global database. 
Although the update language encodes propagation of updates on the integrated database
to every database of the data controllers at once, such program can be considered 
as a composition of all LAVs.
Suppose a new company (ID 3) joined the alliance.
Then what the programmer needs to do is to add new 
part in the program of data integrator in a modular way to describe how updates to the
rows exported from company 3 is propagated to the view of the company ($V_3$),
and how such updates are propagated to the source ($S_3$). 
Parts of the integrator program for other companies, and other controller programs for
companies 1 and 2 can remain intact, enjoying the benefit of LAV.
%
Requirements here include ability to suppress
creating materialized view as much as possible to prevent exporting sensitive
data. We will discuss challenges here in Section~\ref{sec:discussion}.

The next section describes how the behaviors described above are
programmed by our update language.


\section{A Bidirectional Language for Describing Update Strategies}
\label{sec:language}


In this section, we explain our BX language for describing (view) update strategies, the core of our programmable architecture, demonstrate how it can be used to program strategies for controlling  and integrating data, and show how the corresponding query can be derived from an update program.

\subsection{Overview of the Language}
\label{sec:RU-overview}

With the traditional approaches, a view definition is given as a query that constructs view tables from source tables.
By contrast, with our approach the programmer writes an update program that takes source tables and (possibly changed) view tables as input, and manipulates them with the aim of putting all information of the view tables into the right places of the source tables.
From this update program we can then automatically derive the corresponding well-behaved query, which is amenable to standard techniques like query optimization and rewriting.

Below we introduce an experimental relational update language, in which the kind of update program described above can be written.
The language follows the now classic combinator-based design~\cite{Lenses, Bohannon:06}, and consists of:
\begin{itemize}
\item atomic instructions (\verb|CHECK| and \verb|UPDATE|) that check the integrity of a view table or overwrite parts of a source table using information from a view table, and
\item composite instructions (\verb|VSPLIT| and \verb|HSPLIT|) that split a view table either vertically or horizontally, and continue to execute further update instructions on the resulting smaller view tables and the source tables.
\end{itemize}
Instead of giving formal definitions, we will illustrate the use of the language with examples (two data controllers in Section~\ref{sec:RU-sharing} and a data integrator in Section~\ref{sec:RU-integration}), after which we will explain how query derivation, the distinguishing feature of the language, is realized (Section~\ref{sec:RU-query-derivation}).

\subsection{Programming Data Controllers}
\label{sec:RU-sharing}

Suppose that a ride-sharing company maintains the following source table about its vehicles:
\begin{flalign*}
~~~
& \texttt{vehicles(\underline{vid}, loc, rid)} &
\end{flalign*}
Recorded for each vehicle are a unique vehicle identifier (\verb|vid|, which is underlined to indicate that it is the primary key of the table), its current location (\verb|loc|), and a request id (\verb|rid|).
For privacy reasons, when sharing vehicle information with the mediator, the company wishes to show only an approximate area where a vehicle is, rather than the precise location.
The company therefore also maintains another source table mapping locations to areas:
\begin{flalign*}
~~~
& \texttt{area\char95map(\underline{loc}, area)} &
\end{flalign*}
The view exposed to the mediator has the following schema:
\begin{flalign*}
~~~
& \texttt{peer1\char95public} & \\[-.75ex]
& \texttt{~~(\underline{vehicle\char95id}, current\char95area, request\char95id)} &
\end{flalign*}
which contains only approximate areas the vehicles are in.
Our task here is to program a data controller that synchronizes this view table and the source tables.

In our approach, to establish the relationship between the source tables (\verb|vehicles| and \verb|area_map|) and the view table (\verb|peer1_public|), we should describe an update strategy, that is, how view information should be used to update the sources.
In particular, with an update strategy we can control what view information can be changed.
For this example, we might allow the mediator to change only \verb|request_id|s but not \verb|vehicle_id|s and \verb|current_area|s.
Our strategy is therefore updating the \verb|rid| attribute of \verb|vehicles| while checking whether other information in the view (vehicle ids and current areas) is intact; if not, we regard the view as invalid and reject the update.

The above strategy is programmed in our language as follows:
\begin{verbatim}
  VSPLIT VIEW peer1_public WITH
    vehicle_id, request_id {
      UPDATE    vid, rid
      IN SOURCE vehicles
      WITH      vehicle_id, request_id
      IN VIEW   peer1_public
    }
    vehicle_id, current_area {
      CHECK VIEW peer1_public EQUALS
        SELECT vid AS vehicle_id,
               area AS current_area
        FROM   vehicles, area_map
        WHERE  vehicles.loc = area_map.loc;
    }
\end{verbatim}
We vertically split (i.e., project) \verb|peer1_public| into two tables, the first one consisting of the two attributes \verb|vehicle_id| and \verb|request_id|, and the second one the two attributes \verb|vehicle_id| and \verb|current_area|.
The two projected tables, still named \verb|peer1_public|, are used respectively in the two parallel updates specified in the blocks enclosed in curly brackets:
\begin{itemize}
\item In the first block, we update \verb|vid| and \verb|rid| in \verb|vehicles| with the corresponding attributes in \verb|peer1_public|.
This is done by matching records in the source and view tables by their keys, i.e., \verb|vid| and \verb|vehicle_id|, replacing \verb|rid| with \verb|request_id|, and keeping the unmentioned attribute \verb|loc| unchanged.
\item In the second block, we make sure that the mediator does not tamper with the attributes \verb|vehicle_id| and \verb|current_area| in \verb|peer1_public| by checking whether \verb|peer1_public| (which, in this block, has only the two attributes) is equal to the result of a query that extracts the source information that should be kept unchanged and translates locations to areas.
The whole update is rejected if this check fails.
\end{itemize}

What is interesting is that from the update program we can automatically extract the corresponding well-behaved query, which is equivalent to:
\begin{verbatim}
  SELECT vid AS vehicle_id, area AS current_area,
         rid AS request_id
  INTO   peer1_public
  FROM   vehicles, area_map
  WHERE  vehicles.loc = area_map.loc;
\end{verbatim}

To show a different data controller, suppose that another company maintains occupied and unoccupied vehicles in two separate tables:
\begin{flalign*}
~~~
& \texttt{occupied\char95vehicles(\underline{vid}, area, rid)} & \\[-.75ex]
& \texttt{unoccupied\char95vehicles(\underline{vid}, area)} &
\end{flalign*}
There is no request id for an unoccupied vehicle, so we omit the \verb|rid| attribute in \verb|unoccupied_vehicles|; also, unlike the first company, this company stores only the approximate areas the vehicles are in.
The view exposed to the mediator is the same as the first company's except for the table name:
\begin{flalign*}
~~~
& \texttt{peer2\char95public} & \\[-.75ex]
& \texttt{~~(\underline{vehicle\char95id}, current\char95area, request\char95id)} &
\end{flalign*}
The update strategy is specified as follows:
\begin{verbatim}
  HSPLIT VIEW peer2_public ON request_id
    null {
      UPDATE    vid, area
      IN SOURCE unoccupied_vehicles
      WITH      vehicle_id, current_area
      IN VIEW   peer2_public
    }
    OTHERWISE {
      UPDATE    vid, area, rid
      IN SOURCE occupied_vehicles
      WITH      vehicle_id, current_area,
                request_id
      IN VIEW   peer2_public
    }
\end{verbatim}
This time we horizontally split (i.e., select) \verb|peer2_public| into two tables based on the \verb|request_id| attribute: the first table consists of all the records whose \verb|request_id| is null, and all other records are collected in the second table.
The two tables (still named \verb|peer2_public|) are then used to update \verb|unoccupied_vehicles| and \verb|occupied_vehicles| respectively.
Again from this program we can derive the corresponding well-behaved query:
\begin{verbatim}
  SELECT *
  INTO   peer2_public
  FROM   SELECT vid AS vehicle_id,
                area AS current_area,
                null AS request_id
         FROM   unoccupied_vehicles
         UNION
         SELECT vid AS vehicle_id,
                area AS current_area,
                rid AS request_id
         FROM   occupied_vehicles;
\end{verbatim}

\subsection{Programming Data Integrators}
\label{sec:RU-integration}

Instead of the two tables \verb|peer1_public| and \verb|peer2_public| provided by the ride-sharing companies, the mediator prefers to work on a single table:
\begin{flalign*}
~~~
\texttt{all\char95vehicles(}&\texttt{\underline{company\char95id}, \underline{vehicle\char95id},} & \\[-.75ex]
&\texttt{current\char95area, request\char95id)} &
\end{flalign*}
Synchronization between the table \verb|all_vehicles| and the two tables \verb|peer1_public| and \verb|peer2_public| is performed by a data integrator, which, like data controllers, can be programmed with our language by specifying how to put \verb|all_vehicles| into \verb|peer1_public| and \verb|peer2_public|.
The program is similar to the one for the second company's data controller: we perform a horizontal split on \verb|company_id| and put the resulting tables into either \verb|peer1_public| or \verb|peer2_public|.
\begin{verbatim}
  HSPLIT VIEW all_vehicles ON company_id
    1 {
      UPDATE    vehicle_id, current_area,
                request_id
      IN SOURCE peer1_public
      WITH      vehicle_id, current_area,
                request_id
      IN VIEW   all_vehicles
    }
    2 {
      UPDATE    vehicle_id, current_area,
                request_id
      IN SOURCE peer2_public
      WITH      vehicle_id, current_area,
                request_id
      IN VIEW   all_vehicles
    }
\end{verbatim}
And the derived query is:
\begin{verbatim}
  SELECT *
  INTO   all_vehicles
  FROM   SELECT *, 1 AS company_id
         FROM   peer1_public
         UNION
         SELECT *, 2 AS company_id
         FROM   peer2_public;
\end{verbatim}

\subsection{Query Derivation}
\label{sec:RU-query-derivation}

\newcommand{\coladj}{\kern-\medskipamount}
\newcommand{\mindent}{\kern.5em}

The precise semantics of the language is intricate and requires careful design (by imposing syntactic and semantic constraints) to guarantee well-behavedness like what Bohannon et~al. did with their ``relational lenses''~\cite{Bohannon:06}, but we will not go into the details here.
(See Section~\ref{sec:relatedWork} for a discussion of Bohannon et~al.'s work.)
Instead, we will only explain the language's distinguishing feature: the mechanism of query derivation.
Each statement --- \verb|CHECK|, \verb|UPDATE|, \verb|VSPLIT|, or \verb|HSPLIT| --- corresponds to a kind of query, and the correspondence, i.e., translation from update programs to queries, can be described syntactically.

The simplest case is a \verb|CHECK| statement:
\begin{flalign*}
\mindent & \begin{array}{l}
\mathtt{CHECK\ VIEW}\ \mathit{viewTable}\ \mathtt{EQUALS}\ \mathit{srcQuery}
\end{array} &
\end{flalign*}
whose corresponding query is exactly $\mathit{srcQuery}$.
A slightly more interesting case is an \verb|UPDATE| statement:
\begin{flalign*}
\mindent
&\begin{array}{l}
\mathtt{UPDATE}\ \mathit{srcAttrs}\ \mathtt{IN\ SOURCE}\ \mathit{srcTable} \\
\mathtt{WITH}\ \mathit{viewAttrs}\ \mathtt{IN\ VIEW}\ \mathit{viewTable}
\end{array} &
\end{flalign*}
which is translated into projection and attribute renaming:
\begin{flalign*}
\mindent & \begin{array}{ll}
\mathtt{SELECT}\coladj & \mathit{srcAttrs}\ \mathtt{AS}\ \mathit{viewAttrs} \\
\mathtt{INTO} & \mathit{viewTable} \\
\mathtt{FROM} & \mathit{srcTable};
\end{array} &
\end{flalign*}

When it comes to composite statements, i.e., \verb|VSPLIT| and \verb|HSPLIT|, the general plan is to derive queries from the blocks and then assemble the results of the queries.
For a \verb|VSPLIT| statement:
\begin{flalign*}
\mindent & \begin{array}{l}
\mathtt{VSPLIT\ VIEW}\ \mathit{viewTable}\ \mathtt{WITH} \\
\quad\mathrlap{\mathit{ViewAttrs}_1}\phantom{\mathit{ViewAttrs}_n}\ \{\mathrlap{p_1}\phantom{p_n}\} \\
\quad\ldots \\
\quad\mathit{ViewAttrs}_n\ \{p_n\}
\end{array} &
\end{flalign*}
we translate it into a join:
\begin{flalign*}
\mindent & \begin{array}{ll}
\multicolumn{2}{l}{\textit{query derived from $p_1$,}} \\
\multicolumn{2}{l}{\quad\textit{where all occurrences of `$\mathit{viewTable}$\!' are replaced with}} \\
\multicolumn{2}{l}{\quad\textit{an unused table name `$\mathit{tmpTable}_1$\!'\,};} \\[1ex]
\multicolumn{2}{l}{\ldots} \\[1ex]
\multicolumn{2}{l}{\textit{query derived from $p_n$,}} \\
\multicolumn{2}{l}{\quad\textit{where all occurrences of `$\mathit{viewTable}$\!' are replaced with}} \\
\multicolumn{2}{l}{\quad\textit{an unused table name `$\mathit{tmpTable}_n$\!'\,};} \\[1ex]
\mathtt{SELECT}\coladj & {\textstyle\bigcup_i \mathit{viewAttrs}_i} \\
\mathtt{INTO} & \mathit{viewTable} \\
\mathtt{FROM} & \mathit{tmpTable}_1, \ldots, \mathit{tmpTable}_n \\
\mathtt{WHERE} & \mathit{tmpTable}_i.\mathit{attr} = \mathit{tmpTable}_j.\mathit{attr} \\
& ~~ \textit{for all}\ i \neq j\ \textit{and}\ \mathit{attr} \in \mathit{viewAttrs}_i \cap \mathit{viewAttrs}_j;
\end{array} &
\end{flalign*}
And for an \verb|HSPLIT| statement:
\begin{flalign*}
\mindent & \begin{array}{l}
\mathtt{HSPLIT\ VIEW}\ \mathit{viewTable}\ \mathtt{ON}\ \mathit{viewAttr} \\
\mathit{value}_1\ \{p_1\} \\
\ldots \\
\mathtt{OTHERWISE}\ \{p_n\}
\end{array} &
\end{flalign*}
we translate it into a union:
\begin{flalign*}
\mindent & \begin{array}{lll}
\multicolumn{3}{l}{\textit{query derived from $p_1$,}} \\
\multicolumn{3}{l}{\quad\textit{where all occurrences of `$\mathit{viewTable}$\!' are replaced with}} \\
\multicolumn{3}{l}{\quad\textit{an unused table name `$\mathit{tmpTable}_1$\!'\,};} \\[1ex]
\multicolumn{3}{l}{\ldots} \\[1ex]
\multicolumn{3}{l}{\textit{query derived from $p_n$,}} \\
\multicolumn{3}{l}{\quad\textit{where all occurrences of `$\mathit{viewTable}$\!' are replaced with}} \\
\multicolumn{3}{l}{\quad\textit{an unused table name `$\mathit{tmpTable}_n$\!'\,};} \\[1ex]
\mathtt{SELECT}\coladj & * \\
\mathtt{INTO} & \multicolumn{2}{l}{\mathit{viewTable}} \\
\mathtt{FROM} & \mathtt{SELECT}\coladj & *, \mathit{value}_1\ \mathtt{AS}\ \mathit{viewAttr} \\
& \mathtt{FROM} & \mathit{tmpTable}_1 \\
& \mathtt{UNION} \\
& \ldots \\
& \mathtt{UNION} \\
& \mathtt{SELECT}\coladj & * \\
& \mathtt{FROM} & \mathit{tmpTable}_n;
\end{array} &
\end{flalign*}

We deliberately keep the language simple so that we can explain its query derivation mechanism in a straightforward manner.
In general, query derivation will be much more complex for more expressive update languages (e.g., BiFluX~\cite{PaZH14}).

%

\section{Implementation in PostgreSQL}
\label{sec:implementation}

In this section, we briefly explain how the programmable
architecture given in Section \ref{sec:architecture} can be
implemented in a conventional DBMS, PostgreSQL, and concretely
demonstrate how the ride-sharing alliance system is actually
implemented with PostgreSQL.

%

\subsection{Basic Ideas}
\label{subsec:basicideas}

As stated in Section~\ref{sec:architecture},
our programmable architecture has three parts, namely,
data controller, data integrator, and data connector.
The basic ideas for implementing them are described below.

\begin{enumerate}
\item
Data controller.
\\
Each data controller uses materialized views
(i.e., $V_1,\ldots,V_n$ in the lower part of Figure~\ref{fig:architecture})
to export a part of its own, original data to others.
The materialized views are derived by the BX (i.e., predefined update
strategy and the corresponding view definition)
established within the data controller. 
In order to control the exported data,
an update on the materialized views is interpreted first by
the update strategy (and hence, by the data provider).
The update is then rewritten and propagated to the source tables
of the views only when the given update is acceptable to the update
strategy.
\item
Data integrator.
\\
The data integrator keeps copies
(i.e., $V_1,\ldots,V_n$ in the upper part of Figure~\ref{fig:architecture})
of the materialized views of data controllers.
The integrated view 
(i.e., $I$ in Figure~\ref{fig:architecture})
is derived and materialized from those copies by BX.
An update on the integrated view is rewritten to updates on the
copies, and then the updates are propagated to the materialized
views at the data controllers.
\item
Data connector.
\\
The copies at the data integrator are synchronized with the materialized
views at the data controllers, and vice versa.
This part is not difficult to implement by utilizing
PostgreSQL's functionalities such as
\emph{Foreign Data Wrappers} (FDW for short).
\end{enumerate}

From the perspective of implementation,
the key parts are data controller and data integrator
because they contain BX inside.
The detail of the implementation,
especially how to realize the query rewriting and propagation by BX,
will be explained in Section~\ref{subsec:impBXinPosgreSQL}.
After that, in Section~\ref{sec:impRideSharing},
it is demonstrated that the three parts can be implemented with PostgreSQL
through the ride-sharing alliance example.


\subsection{Implementing BX in PostgreSQL}
\label{subsec:impBXinPosgreSQL}
In this subsection, we briefly describes an important implementation issue that 
how the BX can be systematically implemented in PostgreSQL. 
We use {\em triggers}, which can be used to define functions 
based on ECA(Event, Condition and Action)-rules. 
As a language for defining trigger functions,
a PostgreSQL-dialect of PL/SQL called PL/pgSQL
has enough expressive power to implement the update strategy.
As described above, there are two kinds of BX 
in the programmable architecture. One is established 
in the data controller, the other is established in the data
integrator. 
For each BX (a derived $get$ and a predefined $put$), bidirectional
update propagation can be realized by 
preparing two triggers, one is defined on views and the other is
defined on tables, which compose the views. 
Note that one may implement a bidirectional update propagation
by preparing two trigger functions, independently. However, in such 
implementations, the important property, the round-trip property shown in
the introduction, can not be guaranteed. 
Instead, we define two trigger functions satisfying the property 
by translating through the well-behaved BX (a $get$ and a $put$). 
Note also that we need to avoid falling in infinite loops by
pulling the triggers in both sides.  


Since both the views of self-controllable data in the data controller
and integrated views in data integrator are materialized, both
an incremental version of view maintenance for 
$get$ and an incremental version of $put$ with respect to the updates are
needed to implement the updates propagation in trigger functions
to reflect both updates on sources and views efficiently.
So, the incremental version of view maintenance for $get$ with respect
to an update on a source specifies how to reflect an update to the
materialized view when the update is executed to source data.
Whereas, the incremental version of $put$ specifies
how to reflect an update to the source data
when the update is executed to a view.
In other words, for given a source database $S$, a view definition $get$
and an update $w$ on the source $S$, an incremental version of view
maintenance for $get$ with respect to $w$ denoted by $w'$
satisfies the following equation: 
\[
get(w(S)) = w' (get(S))
\]
Also, for given a materialized view $V$, an update strategy $put$ and an
update $u$ on the view $V$, an incremental version of $put$ with
respect to $u$ denoted by $u'$ satisfies the following equation:
\[
put(S, u(V)) = u'(V)
\]
Note that we can apply the existing work \cite{AKKN12} and
\cite{Horn17} to obtain an incremental view maintenance for $get$ and 
an incremental update strategy for $put$, respectively. 

Now, we show, for given a BX (a $get$ and a $put$), the whole steps to
implement two trigger functions, one is for a $get$ and the other is
for a $put$ by using the examples shown in Section~\ref{sec:RU-sharing}. 
For implementing a $get$, we use the following two steps:
\begin{enumerate}
\item[(g1)] Deriving an incremental version of view maintenance. \\
An incremental version of view maintenance for $get$ with respect to an update on
a source can be obtained based on a static analysis \cite{AKKN12}.  
For example, when the following insertion $w$ is
executed to \verb|vehicles| shown in Section~\ref{sec:RU-sharing}: 
\begin{verbatim}
  INSERT INTO vehicles
  VALUES (new_vid, new_loc, new_rid)
\end{verbatim}
the following $w'$ can be obtained from the view definition 
$get$ derived from the update strategy for \verb|peer1_public| shown in 
Section~\ref{sec:RU-sharing}: 
\begin{verbatim}
  INSERT INTO public_peer1
  SELECT new_vid, area, new_rid
  FROM   area_map
  WHERE  area_map.loc = new_loc
\end{verbatim}
\item[(g2)] Translating the incremental view maintenance into a trigger function.\\
  For example, the above incremental version of view maintenance can be translated
  into the following pseudo code in trigger function defined on
  \verb|vehicles|: 
\begin{verbatim}
  CREATE OR REPLACE FUNCTION
    vehicles()
  RETURNS trigger AS $$
    BEGIN
      IF TG_OP ='INSERT' THEN
        INSERT INTO public_peer1
        SELECT NEW.vid, area, NEW.rid
        FROM area_map
        WHERE area_map.loc=New.loc;
      ELSE ...
      END IF;
      RETURN NEW;
    END;
  $$ LANGUAGE plpgsql;
\end{verbatim}
The above trigger function works that when an event of insertion to
\verb|vehicles| is happened, the insertion of the incremental view
maintenance to \verb|public_peer1| is also executed. 
\end{enumerate}  

Similarly, for implementing a $put$, we use the following two steps:
\begin{enumerate}
\item[(p1)] Deriving an incremental version of an update strategy.\\
An incremental version of $put$ with respect to an update on a view can
be obtained based on a static analysis \cite{Horn17}. 
For example, when the following update $u$ is executed to
\verb|peer1_public|: 
\begin{verbatim}
  UPDATE peer1_public
  SET    request_id = new_id
  WHERE  viehicle_id = id
\end{verbatim}
the following $u'$ can be
obtained for a given update strategy shown in
Section~\ref{sec:RU-sharing}: 
\begin{verbatim}
  UPDATE vehicle
  SET    rid = new_rid
  WHERE  vid = id
\end{verbatim}
\item[(p2)] Translating the incremental $put$ into a trigger function.\\
  For example, the above incremental version of update strategy can be
  translated into the following
  pseudo code in trigger function is defined on \verb|peer1_public|:
\begin{verbatim}
  CREATE OR REPLACE FUNCTION
  update_peer1_public()
  RETURNS trigger AS $$
    BEGIN
      IF TG_OP = 'UPDATE' THEN
        UPDATE vehicle
        SET
          rid = NEW.request_id
        WHERE
          vid = NEW.vehicle_id;
      ELSE ...
      END IF;
      RETURN NEW;
    END;
  $$ LANGUAGE plpgsql;
\end{verbatim}
The above trigger function works that when an event of update on 
\verb|public_peer1| is happened, the update of the incremental $put$ 
on \verb|vehicles| is also executed. 
\end{enumerate}

\subsection{Example: Implementation of Privacy-Preserving Ride-Sharing Alliance System}
\label{sec:impRideSharing}
\begin{figure*}[ttt]
  \centerline{\includegraphics[width=0.7\textwidth]{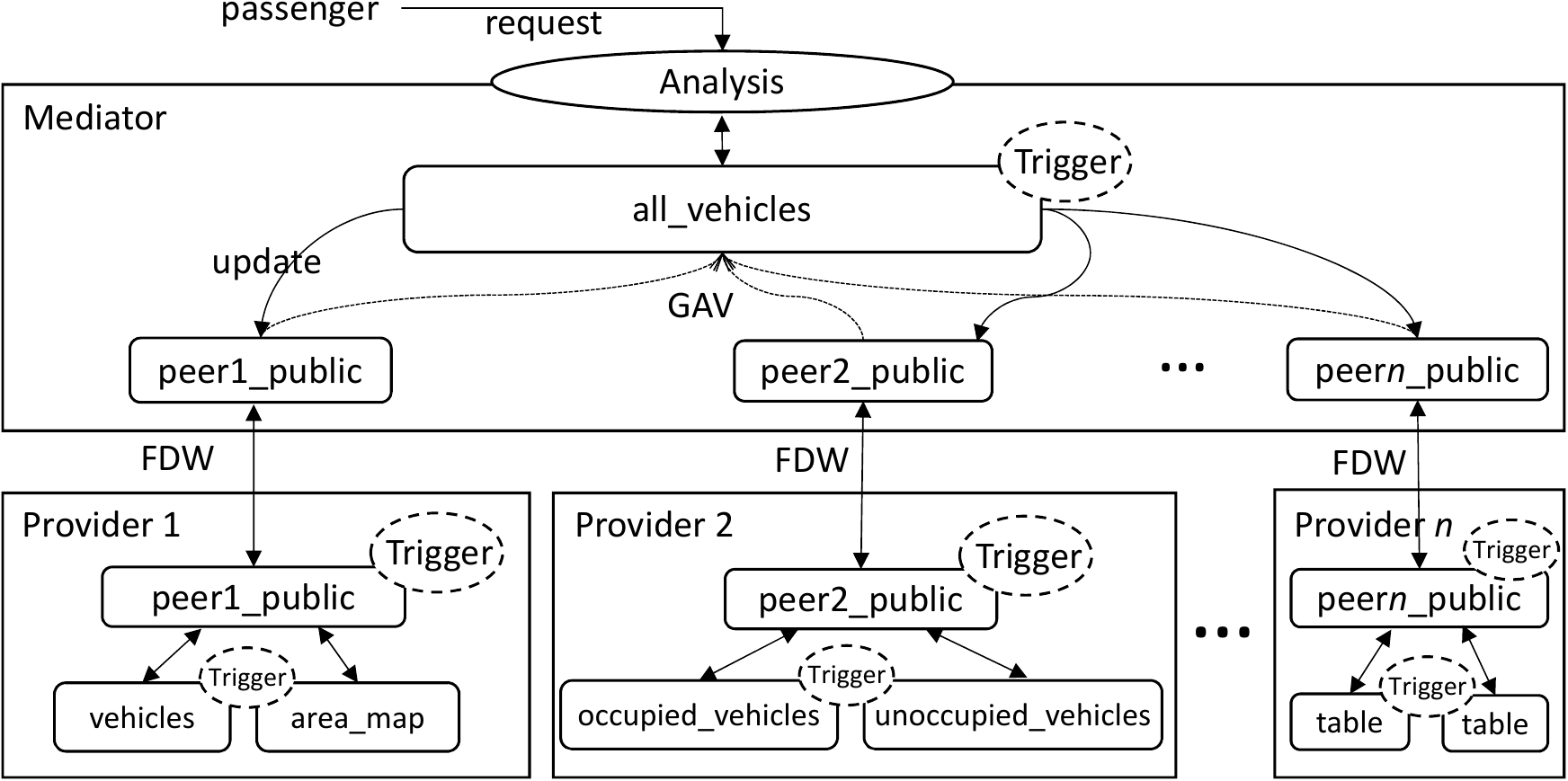}}
  \caption{A framework for Privacy-Preserving Ride-Sharing Alliance System}
  \label{fig:rideSharingSystem}
\end{figure*}
We explain the implementation details of the ride-sharing alliance system as described in Section \ref{sec:ridesharing}.
The system provides a data to book an appropriate taxi for a passenger.
Figure \ref{fig:rideSharingSystem} shows a framework for the system.
The mediator integrates data from multiple taxi companies that manage their own databases whose schemes are different each other.


\subsubsection{Implementing Basic Parts}
To implement the framework of the system, we need to implement the basic parts (i.e., data controller, data integrator, and data connector in Section \ref{subsec:basicideas}).
In addition, we implement a function to analysis user requests.

\subsubsection*{Data controller}
Each taxi company maintains their own database.
To simplify the discussion, we assume two taxi companies use different schema as shown in Section \ref{sec:language}. 
In each database, there are multiple tables and one materialized view ({\tt peer$i$\_public}). 
{\tt peer$i$\_public} is used to provide information from each company to the mediator. 
This view conceals the private information such as the precise locations of the vehicles.  

Each database of a taxi company is {\em self-controlable} since it is managed only by the taxi company while {\tt peer$i$\_public} is used for exporting data to the mediator.
Each table (such as {\tt vehicles} table at provider 1) can be updated from outside only through the update of {\tt peer$i$\_public}.
To this end, we use  triggers provided described in Section \ref{subsec:impBXinPosgreSQL}.
This trigger is executed at each provider (the lower part of Figure \ref{fig:rideSharingSystem}).

\subsubsection*{Data integrator}
The mediator integrates the disclosed information on each peer as an integrated view {\tt all$\_$vehicles}, and it provides data in a common schema for the analysis. 
{\tt all$\_$vehicles} is created  as a virtual view on the mediator by the query descried in Section \ref{sec:RU-integration}.
Since the mediator updates {\tt all$\_$vehicles} for booking taxis, the update is propagated to the database at each peer.
We have two triggers on mediator for updating the {\tt all$\_$vehicles} view and  {\tt peer$i$$\_$public} view of each provider $i$.
The triggers for updating {\tt all$\_$vehicles} from provider 1 and  {\tt peer1$\_$public} from the mediator are the following.

%

\begin{quote}
\begin{verbatim}
CREATE OR REPLACE FUNCTION 
 update_peer1_public_mediator()
RETURNS TRIGGER
AS $$
 BEGIN
    IF NEW.company_id = 1 THEN
      UPDATE peer1_public 
      SET
        request_id=NEW.request_id
      WHERE vehicle_id=NEW.vehicle_id;
    ELSIF NEW.company_id = 2 THEN
    ...
    RETURN NEW;
  END;
$$;
\end{verbatim}
\end{quote}

\begin{quote}
\begin{verbatim}
CREATE OR REPLACE FUNCTION 
 update_all_vehicleson_on_peer1()
RETURNS TRIGGER
AS $$
 BEGIN
      IF TG_OP = 'UPDATE' THEN
        UPDATE all_vehicles 
        SET
          1 AS company_id
          request_id=NEW.req_id
          current_area = NEW.current_area
        WHERE
          vehicle_id = NEW.vehicle_id
        END IF;
    RETURN NEW;
  END;
$$;
\end{verbatim}
\end{quote}
Both triggers are executed at the mediator (the upper part of Figure \ref{fig:rideSharingSystem}).

\subsubsection*{Data connector}
The mediator and each provider are connected by FDW, and thus each {\tt peer$i$$\_$public} on the mediator is defined as a foreign table.
The databases on the providers may not allow updating because a vehicle that is tried to be booked may be already booked by other passengers.
For handling this case, each peer sends back error messages, and if the mediator receives the messages, it analyzes the passengers' requests again.
\subsubsection{Data Analysis on Mediator}
\label{subsec:analysis}

The system analyzes the passenger request, and then shows a list of taxis to passengers.
We describe how the mediator analyzes data on the integrated view. 
The system analyzes the benefits of taxi companies based on the time of the request and the locations of the start point, destination, and vehicles. 
For example, it lists $K$ taxis with the highest benefits.
A pseudocode  to list $K$ taxis is as follows:
\begin{quote}
\begin{verbatim}
FUNCTION candidate_taxis (
   u_pickup_location  int,
   ...
   )
RETURNS TABLE (
  u_company_id    int,
  u_vehicle_id   int,
  total_benefit  int
 ...
  )
 ...
  ORDER BY
    total_benefit DESC
  LIMIT K;
\end{verbatim}
\end{quote}

\section{Discussion}
\label{sec:discussion}

In this section, we discuss important issues in our proposed programmable architecture, including the application scope (through two more interesting application examples), challenges in practical design and implementation of the architecture, and the evaluation criteria.

\subsection{Potential Applications}

\subsubsection{Personal Data Market}

A platform for personal data market
is one of important potential applications of 
our 
programmable
architecture (Fig.~\ref{fig:architecture}.)
The importance of personal data market for general \cite{oecd_exploring_2013}\cite{ferretti_federico_eu_nodate} and
specific (e.g. geosocial \cite{yaron_kanza_online_2015}) data
is widely acknowledged.
A number of related works has been emerging which include
princing \cite{muschalle_pricing_2012}\cite{roth_technical_2017}\cite{DBLP:journals/tods/LiLMS14,li_theory_2017}\cite{li_first_2017},
auction \cite{roth2012buying}\cite{DBLP:journals/geb/GhoshR15a}
and architecture \cite{montjoye_openpds:_2014}.

To protect privacy, 
data owners should retain control over their own personal data
and should have a right to specify a subset of data to be
shipped to a marketplace.
This scenario of personal data onwers and marketplace 
nicely fits with our programmable architecture
shown in Fig.~\ref{fig:architecture}.
In this scenario of personal data market,
data controllers and data integrator in Fig.~\ref{fig:architecture}
are regarded as data owners and marketplace, respectively.
Data owners release a subset of their data as views
$V_1, \dots, V_n$. The marketmaker sells 
an integrated view $I$ to 
the buyer (which is represented as a user in Fig.~\ref{fig:architecture}.)

Current studies in personal data market assume simple transactions
without negotiation between data owners and buyers.
However, we forsee transactions in practical
personal data market are not straighforward in
many cases because of its inherent complexity. 
The price which data owners ask is not a simple numeric value but might
be a fuction of degree of privacy disclosure (which, among others,
is measured by $\epsilon$ of differential privacy.)
Meanwhile, buyers may offer bonus to data owners as an incentive to
release personal data.
We expect the BX mechanism 
in our archecture plays an important role to support such complex
negotiation between data owners and buyers.

\subsubsection{Management of Scientific Metadata}

For the management of scientific data, 
the metadata 
plays an important role. 
%
%
In the domain of earth science, we generally make dataset-level metadata which describe 
the overview of the dataset such as dataset name, dataset creator, abstract text, 
spatiotemporal information, keywords, and so on. 
Such {\it scientific metadata} are in different formats depending on the target fields, and 
managed at each organization the data belong to. 
Therefore, we have many local databases for metadata management, 
which have their own update strategy and security policy.

With the growth of interdisciplinary data science, 
there are some systems for integrated management of scientific data and metadata. 
The systems like GCMD\footnote{https://gcmd.nasa.gov/} and 
PANGAEA\footnote{https://www.pangaea.de/} accept scientific metadata 
of various earth science fields in their specified format, 
while the search systems of GEOSS \cite{geoss2013} and DIAS \cite{dias2017} 
work as mediators to integrate scientific metadata from local databases. 
Note that the focus of these systems is providing search functions, and 
therefore analysis and updates on the integrated data are not well considered. 
Our architecture can be naturally applied to the management of scientific metadata, and 
for example, it will be useful for the curation on the integrated data.

\subsection{Future Challenges}

In the following, we highlight some important challenges in design and implementation of the programmable architecture, and show possible ways to tackle them.

\subsubsection{Issues in Bidirectional Update Languages}

The bidirectional relational update language in Section \ref{sec:language} serves to demonstrate possibility of designing a language to specify (well-behaved) view update strategies, from which the unique corresponding query can be automatically derived.  This kind of framework has been studied in other settings for some years and recently established more convincingly as a plausible approach \cite{KoHu18}, but it has not been instantiated for relational databases. The future  challenges on this language are as follows.

\subsubsection*{Language extension}

The current language is rudimentary and not powerful enough for more sophisticated scenarios. Despite its simplicity, what underlies the language is a general-purpose update programming model (as opposed to a restrictive model that translates view modifications to source modifications using a hard-wired translation logic), and the language can eventually be extended with more programming constructs so that programmers can freely and fully customize their update strategies.

\subsubsection*{Well-behavedness of update strategies}

Certainly not any view update strategy is well-behaved in the sense that there exists a query that can be paired with it to form a bidirectional transformation. We omit the discussion about how to validate the well-behaveness of an update strategy in our bidirectional relational update language, but as discussed in Section \ref{sec:putbx}, we can follow the idea in \cite{HuPS14} to design a static analysis algorithm to do this validation.

\subsubsection*{Efficient incrementalization of view-updating}

Challenges in propagating updates through relational views include
avoiding to use materialized views as much as possible.
In current state of the art of relational lenses,
although Bohannon et al.,'s approach\cite{Bohannon:08} has been incrementalized by
Horn and Cheney\cite{Horn17,HornCheney17}, their approach still requires querying source data
to compute update translation. This is because some of the updates
on the view may affect the original source database indirectly
through functional dependency.
We could reduce such query to source database by implementing
query in a compositional way, and make a component query closer
to the database selective enough so that backward execution of
the subsequent queries may access only limited materialized view.
In other words, we believe that the known push-down optimization can be considered
not only for view computation but also for view updating.

\subsubsection{Implementation Issues}

We roughly show that the programmable architecture can be implemented over PostgreSQL in Section \ref{sec:implementation}, and the future challenges are as follows.

\subsubsection*{Automatic translation}
The core part of this translation from the higher-level description to the lower-level implementation over PostgreSQL is the translation of update strategies to a pair of triggers for propagating updates bidirectionally. This includes a systematic derivation queries from update strategies and an automatic incrementalization of both of them. Although theoretically all of them can be done, but it is a good engineering work in practice to implement the above efficiently.

\subsubsection*{FDW and trigger functions}
Since we rely on PostgreSQL FDW and trigger functions for data synchronization among database servers, we share their benefits and limitations.
In particular, the transaction is supported across/inside database servers~\footnote{https://www.postgresql.org/docs/10/static/postgres-fdw.html}.
FDW supports nested transactions between local server and remote servers, so commits and aborts are synchronous between them. The triggers are designed to be executed in the same transaction of its main transaction. 
However, a major limitation is that distributed deadlock cannot be detected by FDW, so the applications that access to the databases are carefully designed to avoid distributed deadlock.

\subsubsection*{Heterogeneous databases}
The current implementation uses PostgreSQL, but we can easily extend our system to use other relational database management systems. We should also support NoSQL databases (Apache HBase, Drill, MongoDB, etc), since they are widely used in various applications. The difficulty of using NoSQL as the data controller is that it does not support logical data model and SQL as a query language. It is our future work to support NoSQL so that our system can be applicable to more heterogeneous environment.

\subsubsection*{View materialization}
Near real-time event streams are becoming a key feature of recent applications.
For example, Twitter and Facebook allow users to create a personalized feed (timeline view) by selecting their friends to follow. 
The timeline view collects the latest posts of the friends in real-time.
The ride-sharing alliance system needs to provide best matches between passengers and taxis based on their latest locations. To achieve efficient real-time services, we need to adaptively materialize views to improve the system performance. A typical approach is to use control table \cite{Zhou2007} that contains hot data (users and/or taxis) to be dynamically materialized. 
Feeding Frenzy~\cite{Silberstein2010} is a framework to selectively materialize users' event feeds in Twitter system.
It is our future work to employ the adaptive view materialization techniques so that the best match can be found efficiently by materializing the area of the taxis in busy area at the data integrator side.

\subsection{Successful Criteria}

The output of the project would be a general software environment supporting people to develop a dependable system for controlling and integrating decentralized data in a systematic and productive way.
To demonstrate the usefulness of the environment, we will show how to use it to develop some concrete systems such as those mentioned in this paper, and we would go even further to construct some specific software environments for productive development of some special but widely used systems such as health-care systems or publishing systems.

%
%
%
%
%
%
%

\section{Related Work}
\label{sec:relatedWork}
\vspace{-1mm}
\vspace{-1mm}
\subsubsection*{Bidirectional transformation}

Relational lens~\cite{Bohannon:06} is a linguistic approach to the view update problem
for relational databases.
It is based on the notion of lenses -- originally
proposed for trees~\cite{Lenses} -- combinators equipped with well-behaved
bidirectional semantics.
In case of relational lenses,
relational operators selection, projection and join are bidirectionalized
while composition of them are achieved by the composition lens
which preserves well-behavedness by a type system of lenses.
A type of a relational lens specifies
a domain (for a set of sources) and range (for a set of views).
%
Types can also represent functional dependencies,
so typing rule entails
manipulating functional dependencies sometimes in non-trivial way, like
decomposition.
%
Bohannon et~al.'s typing rules are highly declarative, for example,
by including judgments of a particular functional dependencies to be satisfied
by any input relation.
Bohannon et~al.'s technical report~\cite{Bohannon05TR} includes
static manipulation of functional dependencies to facilitate such judgments.
However, there are constraints imposed by Bohannon et~al.'s approach and
its programmability.
The join lens
requires that the join key
functionally determines the entire attributes of the right relation.
Additionally, predicates for the source relations should
be independent from
the output parts of the functional dependencies.
As for programmability of Bohannon el~al.'s approach,
they can control
how deletions of a tuple in the view
reflect to deletions of a tuple in the source.
However, it cannot depend on the source relation like our approach
but just refers deleted tuples in the view.

Horn and Cheney~\cite{Horn17,HornCheney17} incrementalized the relational
lenses by providing their own putback semantics that takes,
instead of states of updated views, set of tuples that are
inserted or deleted in the views (called deltas) where modifications
are represented by a pair of insertion of new tuple after modification
and deletion of old tuple before modification.
Their representation of updates
are thus compact. They translate these deltas to source
through compositions of lenses.
Bohannon et al.'s updates are made
explicit to achieve this translation, by per-tuple representations
associated with functional dependencies needed for their adjustment,
and manipulable representation of predicates in which
occurrences of attribute names are replaced by an expression
that would calculate its updated value.
That achieves efficient
representation of adjustments of attributes.
%
Significant speedup over state-based
approach has been observed by the authors.
Since their approach is a natural extension of Bohannon et al.'s,
the adjustment of source tuples violating functional dependencies
requires access to source.
However, different from our general update language, they focus on
incremental semantics for the specific putback functions that are predefined in the basic relational lenses.
We should generalize their incrementalization approach to deal with more general putback functions in   our update language.



%

One attempt has been made to design Brul \cite{ZLKH16}, a putback-based Haskell library for bidirectional transformations on relations. It provides basic combinators for writing the $put$ function with flexible update strategies easily. Unlike our bidirectional update language, Brul is a library which is not as general as ours. In addition, it shows how a $get$ semantics can be given to a $put$ program, but it does not show how an explicit definition of $get$ can be obtained.
\vspace{-1mm}
\subsubsection*{View updating}

To resolve the ambiguity problem in the view updating, Masunaga recently introduced the intention-based approach \cite{Masunaga:2017:IAU:3022227.3022239,Masunaga18IMCOM}.
It shows that the user's view update intention (update strategy) sometimes can be guessed by checking the extension of each view update transformation candidate, which is calculated using temporarily materialized views. Under the intention-based approach, join views and Cartesian product views became updatable in certain cases. However, this "guess" does not guarantee that a unique update intention can be obtained in general. We tackle this problem by let people write their intention explicitly.
\vspace{-1mm}
\subsubsection*{Data integration}

The classical architecture of data integration is \emph{centralized}.
That is, one mediator gathers all the distributed data,
transforms the data according to the schema mappings,
and provides the uniform data to its users.
On the other hand,
\emph{decentralized} data integration, or peer-to-peer data integration,
has been focused on and many prototype systems have been developed
since the beginning of this century.

Piazza~\cite{Halevy2003,Halevy2004} is one of the first projects on
decentralized data integration.
The Piazza system is for integrating distributed XML documents
without using global ontologies.
It provides query answering functionality based on the certain answer
semantics by rewriting given query.
Updating XML documents on peers is out of the scope.

\textsc{Orchestra}~\cite{Ives2005,Karvounarakis2013} is
a successor project of Piazza.
This project is motivated by the need for collaborative sharing of
scientific data, which are produced by independent researchers without
any global agreement.
The novel concept proposed by the project is referred to as
\emph{collaborative data sharing systems} (CDSS for short).
In CDSS, every peer can independently import other peers' data,
modify the imported data,
merge the modified data with its original data,
and then publish the merged data to other peers.
A peer can update the data published by itself.
Such an update is propagated to updates on
its original data and the imported data.
Then, the new published data are imported again by other peers.
Hence, the view update problem between different peers is out of the
scope of this project.
Moreover, in CDSS,
data inconsistency between different peers is positively
allowed because of the motivation, and therefore,
transaction processing over different peers is not realized.

The Hyperion project~\cite{Kementsietsidis2003,Arenas2003}
proposes an architecture of a
\emph{peer database management system} (PDBMS for short).
A PDBMS consists of three components:
an interface to the users, an ordinary DBMS,
and a P2P layer, which is the key component of a PDBMS.
A P2P layer has the following three functionalities:
managing neighbor peer relationship,
query rewriting for answering queries, and
enforcing data consistency upon different peers.
Because successive query rewriting loses
information of the original query,
a framework called GrouPeer~\cite{Kantere2009,Kantere2011}
was proposed to improve the quality of answering queries.
GrouPeer finds semantically similar peers and
makes them neighbors to avoid successive query rewriting.
In these frameworks,
it is unclear whether updating data on other peers is possible.
Even if this is the case,
controlling update strategy does not seem to be supported.

\vspace{-1mm}
\subsubsection*{Ride-sharing}
Research on ride-sharing has a long history.
Refer to comprehensive surveys of ride-sharing for more details \cite{furuhata2013ridesharing,agatz2012optimization}.
On the other hand, there are a few researches about privacy-preserving ride-sharing.
A{\"i}vodji et al.~\cite{aivodji2016privacy} proposed a method for computing pick-up and drop-off points  for a driver and a passenger without disclosing their precise locations by employing a homomorphism encryption.
Goel et al.~\cite{goel2016privacy} proposed a method for matching drivers and passengers while reducing the number of drivers and passengers who know the precise locations of others by approximating location information.
They also employ a review system in order to eliminate malicious drivers or passengers instead of preventing their attacks directly.
Tong et al.~\cite{tong2017privacy} proposed a method utilizing differential privacy which have been known as a powerful technique for protecting privacy recently.
These methods do not consider a ride-sharing alliance which is dealt with by our work.


\section{Conclusion}
\label{sec:conclusion}

In this vision paper, we have proposed a novel perspective of views, which are defined using view update strategies rather than queries.
This perspective stems from the studies of bidirectional transformations within the programming language community, in particular the insight that well-behaved queries are uniquely determined by, and can be derived from, view update strategies.
Based on this insight, we have designed, for relational databases, a new language for defining views with update strategies, from which the corresponding view queries can be automatically derived.
With this relational update language as the core, we have presented a new view-based programmable architecture for data sharing, integration, and analysis.
Within this architecture, system designers can describe the appropriate data sharing and integrating policies by programming them in the relational update language, rather than having to rely on some hard-wired and often limited view-updating logic derived from query-based view definitions.
As an initial validation of the approach, we have implemented a prototype of a ride-sharing alliance system following the architecture.
We believe that it is worth reporting as early as possible the new perspective of views and the view-based programmable data management architecture arising from the new perspective, so that researchers in databases and programming languages can start working together to explore this promising direction.

\balance


\bibliographystyle{abbrv}
\bibliography{references}  


%
%
%
%

\end{document}